# Probing Silicon Carbide with Phase-Modulated Femtosecond Laser Pulses: Insights into Multiphoton Photocurrent

Ahsan Ali[1,2], Chuanliang Wang[1,2], Jinyang Cai[1,2], and Khadga Jung Karki[1,2,3*]

[1] Department of Physics, Guangdong Technion Israel Institute of Technology, Shantou 515063 Guangdong, China; [2]Guangdong Provincial Key Laboratory of Materials and Technologies for Energy Conversion, Guangdong Technion Israel Institute of Technology, Shantou 515063 Guangdong, China;[3]Technion – Israel Institute of Technology, Haifa, 32000, Israel

*Correspondence: khadga.karki@gtiit.edu.cn

**Abstract**

Wide bandgap semiconductors are widely used in photonic technologies due to their advantageous features, such as large optical bandgap, low losses, and fast operational speeds. Silicon carbide is a prototypical wide bandgap semiconductor with high optical nonlinearities, large electron transport, and a high breakdown threshold. Integration of silicon carbide in nonlinear photonics requires a systematic analysis of the multiphoton contribution to the device functionality. Here, multiphoton photocurrent in a silicon carbide photodetector is investigated using phase-modulated femtosecond pulses. Multiphoton absorption is quantified using a 1030 nm phase-modulated pulsed laser. Our measurements show that although the bandgap is less than the energy of three photons, only four-photon absorption has a significant contribution to the photocurrent. We interpret the four-photon absorption as a direct transition from the valance to the conduction band at the Γ point. More importantly, silicon carbide withstands higher excitation intensities compared to other wide bandgap semiconductors making it an ideal system for high-power nonlinear applications.

Keywords: Silicon carbide, Nonlinear response, Intensity modulation, Multiphoton absorption,

**Introduction**

Development in optoelectronics has a tremendous impact on technologies we use in daily life as well as in specialized equipment[1,2]. Devices such as light-emitting diodes (LEDs), photodetectors, and solar cells have become indispensable[3–5]. In optoelectronics, photons are regarded as a primary energy source and carriers of information. Characteristics like high-speed transmission, broadband response, and low propagation losses are essential for applications in telecommunication, lasers, medical imaging, quantum computing, and astronomy[6]. Advanced photo-integrated circuits (PICs) with the desired scalability require suitable material platforms that are efficient in photonic confinement and stable electronic structure[7,8]. Wide bandgap semiconductors such as gallium composites with nitrides, phosphides, and chalcogenides are the common materials integrated to reduce losses with enhanced efficacy[9–14].

In this category, silicon carbide (SiC) is a comparatively new but promising candidate due to its physical and optical properties[15–17]. Optically, it provides a wide transparent window, high refractive index, and nonlinearities, while physically, it has high thermal conductivity, electron saturation velocity, and breakdown electric field. It is a potential candidate to displace Silicon (Si) in high-power and high-frequency applications[18,19]. Similarly, SiC has useful nonlinear optical properties[20,21], such as harmonic generation, phase modulation[22], four-wave mixing, and multiphoton absorption (MPA)[23,24], which are crucial for optoelectronic devices[25,26].

MPA is fundamental to understanding wide bandgap material-based devices' ultrafast intrinsic processes[24]. These processes are observed when semiconductors are excited by ultrashort pulses of laser with intensities in the range of $10^6 - 10^{12}$ W/cm$^2$ (MW/cm$^2$ to TW/cm$^2$) such that the probability of multiple photons interacting with the system simultaneously is significant. The order of multiphoton absorption depends on the material's electronic structure. For example, it has been found that one, two, and three-photon absorptions contribute to photocurrent in gallium phosphide (GaP)-based photodetector while only three-photon absorption contributes in indium gallium nitride (InGaN) photodetectors[27].

Similar to GaP, SiC is an indirect bandgap semiconductor with a wider bandgap. Importantly, SiC is more stable to heat and radiation suggesting it could be a better material for applications in nonlinear optoelectronics. Anticipating its wider use in the near future, we have investigated multiphoton photocurrent in 4H-SiC photodetector using a highly sensitive technique based on phase-modulation of a pair femtosecond pulses with a central wavelength at 1030 nm (photon energy of 1.2 eV). The bandgap of 4H-SiC is 3.26 eV, which is less than three times the photon energy in our measurements. Just based on the bandgap, one expects a three-photon photocurrent. However, only a four-photon photocurrent is observed in SiC at the excitation intensities of 70 to 200 GW/cm$^2$. These results differ significantly from GaP. We discuss the reason why three-photon absorption is absent in SiC and its implications in nonlinear optoelectronics.

## Materials and Methods

A Mach Zehnder interferometer-based experimental setup is used to produce intensity-modulated femtosecond beams[28–30]; the schematic is shown in Fig 1(a). The laser used in this experiment is the Impulse fiber laser from Clark-MXR, Inc., with a wavelength of 1030 nm (1.2 eV photon energy), a pulse width of 300 fs, and a repetition rate of 0.2–25 MHz. A beam splitter divides the single beam into two, passing through the interferometer's arms separately. The carrier frequency $\upsilon$ of each beam is shifted independently using acousto-optic frequency shifters (AOFS) to $\upsilon_1 = \upsilon + p_1$ and $\upsilon_2 = \upsilon + p_2$, where $p_1$ and $p_2$ correspond to acoustic frequencies. The two beams of different frequencies are aligned carefully and combined at the output to produce a beam whose intensity modulates at the resultant frequency of $\Delta\upsilon = p_2 - p_1$. The optimal output shows complete interference between two phase-modulated beams originating from the same source. The expression for the output beam is given as

$$I(t) \propto I_0 (1 + \cos(2\pi\Delta\upsilon t)), \qquad (1)$$

where I(t) is the time-dependent intensity, and $I_0$ is the amplitude. The output of the interferometer is a train of intensity modulated femtosecond pulses as shown in Fig 1(b). The modulation frequency $\Delta\upsilon$ is set to 1kHz, which is substantially smaller than the bandwidth of the photodetector. A parabolic mirror with a focal length of 33 mm is used to focus the beam onto the SiC photodetector (from GaNo Optoelectronics Inc.). The focus spot size is ~20μm. The focus spot can be visually identified as the spot with green emission from the second harmonic generation (SHG)

(see Fig. 1(c)). A preamplifier is used for current-to-voltage conversion, after which the voltage signal is digitized by a 24-bit data acquisition card at a rate of 192,000 samples per second. The SHG emission is also detected using an avalanche photodiode to further characterize the nonlinear response.

**Results and Discussion**

A linear system when perturbed by a sinusoidal stimulus has a response at the same frequency as the perturbing stimulus. Thus, in the case of a photodetector that can be excited by single-photon absorption, one expects photocurrent to modulate at $\Delta v$. We have used a Si photodetector as a reference to check the fidelity of linear response in our setup. The bandgap of Si (1.12 eV) is less than the photon energy at 1030 nm. As expected, the photocurrent response of Si photodiode presented in Fig 1(d) shows response only at $\Delta v = 1$ kHz indicating that the optical setup itself does not impart nonlinearities. Any spurious signals at harmonics of $\Delta v$ are less than two orders of magnitude smaller.

If different orders of multiphoton absorptions contribute to photocurrent, the measured signal as a function of intensity is generally given by:

$$S \propto \alpha \cdot I + \beta \cdot I^2 + \gamma \cdot I^3 + \chi \cdot I^4 ..., \tag{2}$$

where the coefficients are related to the absorption cross-sections of different orders of interactions. In particular, $\alpha$, $\beta$, $\gamma$, and $\chi$ represent the one-photon (1PA), two-photon (2PA), three-photon (3PA), and four-photon (4PA) absorption cross-sections, respectively. In Eq. (2), we have only considered perturbative description of light-matter interaction and neglected any nonperturbative contributions. Usually, nonperturbative interactions become important at intensities $>10^{13}$ Wcm$^{-2}$, which is about two order of magnitude larger than the intensities we have used in our measurements.

Fig 1(e) shows the Fourier transforms of photocurrent signals from the SiC photodetector at different excitation intensities. Distinct modulations are observed at the frequencies $\Delta v$, $2\Delta v$, $3\Delta v$, and $4\Delta v$ in the measurements done with excitation intensities in the range of $7 \times 10^{10}$ W cm$^{-2}$ to $1.99 \times 10^{11}$ W cm$^{-2}$. The highest modulation frequency observed in the signal is, $4\Delta v$, which indicates that contribution from 4PA is detectable at the most in our measurements. It has been shown that in some systems, different orders of multiphoton absorption can contribute simultaneously[27,31,32]. Is it also the case in SiC?

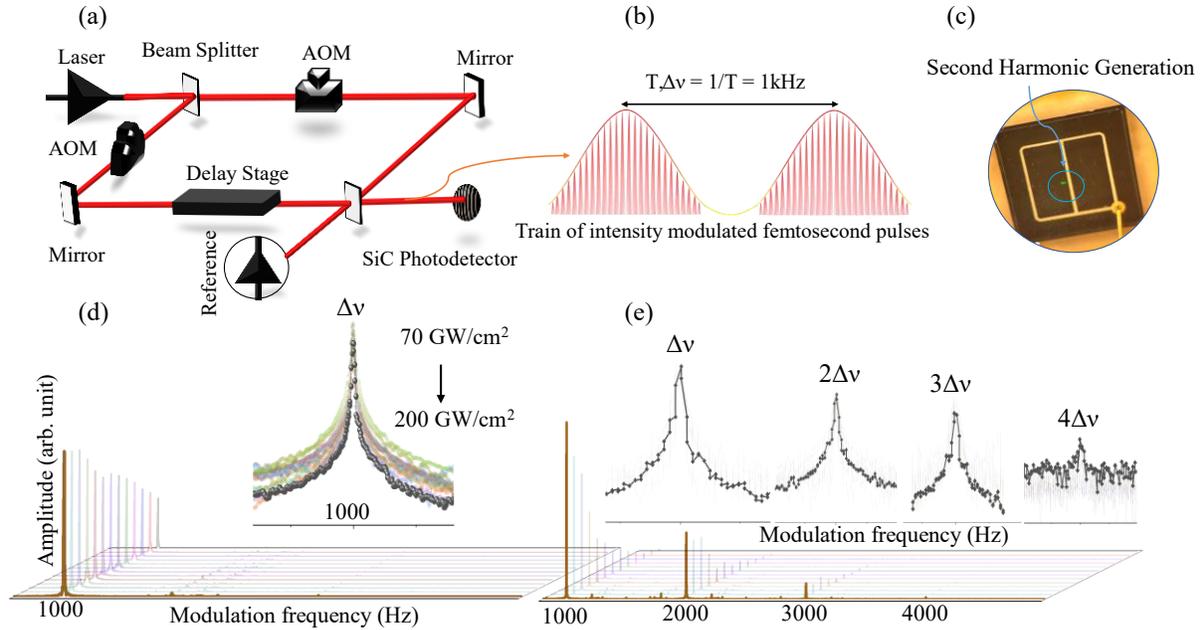

Figure 1. (a) Schematic of experimental design for the modulation of intensity at a single frequency by splitting two beams from the same source (central wavelength 1030 nm) using a Mach-Zehnder interferometer. (b) An illustration of a train of intensity-modulated femtosecond pulses with a modulation frequency of 1kHz. (c) A microscopic image of SiC photodetector with a visible spot of the second harmonic at 515 nm. (d) Response signals of Si photodetector at different powers showing linearity of the system. (e) Nonlinear photocurrent signals from SiC photodetector at different powers, inset shows the peaks at 1,2,3 and 4 kHz.

To answer the question, we analyze the intensity dependence of the signals modulated at $\Delta\nu$, $2\Delta\nu$, $3\Delta\nu$, and $4\Delta\nu$. The measured values of the signals are shown in the log-log plot in Fig 2(a). The linear fits of the data have the slopes of $4.0 \pm 0.1$, $4.0 \pm 0.1$, $4.3 \pm 0.3$, $4.4\pm0.3$ for the signals at the frequencies $\Delta\nu$, $2\Delta\nu$, $3\Delta\nu$, and $4\Delta\nu$, respectively. The quartic dependence of the signals on the intensity at all modulation frequencies indicates that the dominant contribution to photocurrent is from 4PA, while lower order multiphoton absorption processes are negligible. This also implies that, unlike in GaP[27], linear absorption by sub-bandgap traps and defects do not contribute to photocurrent in SiC when excited at 1030 nm. Measurement at a particular spectral window, however, does not ascertain the absence of sub-bandgap absorptions. To test the if a semiconductor is suitable for applications in UV and multiphoton photodetection, it is important to check the photocurrent response at different spectral regions below the bandgap. Thus, we have carried out additional measurements at 515 nm using the second harmonic of 1030 nm. Here, photocurrent is modulated only at Δυ and 2Δυ, which is expected for a 2PA process. Intensity dependence of the signals are shown in Fig 3(a). Linear fits of the signals in the log-log plot have the slopes of 2.0 ± 0.1 for Δυ and 1.9 ± 0.1 for 2Δυ, confirming the prevalence of 2PA. Notably, the contribution from 1PA is not present, providing further evidence to the lack of photocurrent due to sub-bandgap absorptions. These findings show that the SiC devices we have measured are ideal for solar blind detection of UV photons and MPA.

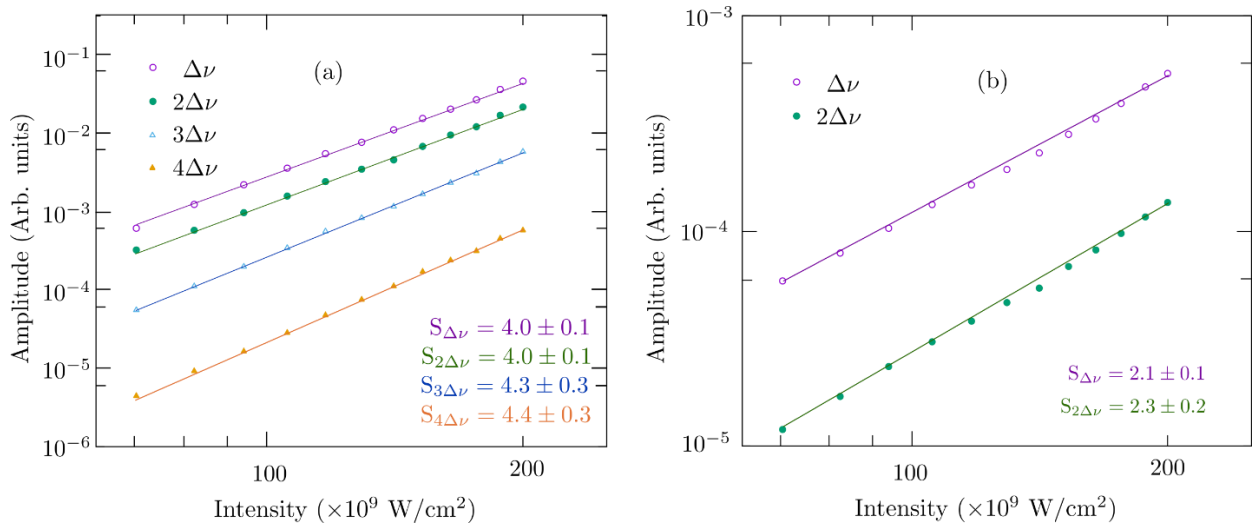

Figure 2. (a) Intensity dependence of photocurrent signals modulated at the frequencies Δυ, 2Δυ, 3Δυ, and 4Δυ, and (b) SHG modulated at the frequencies Δυ, 2Δυ from the SiC photodetector.

Apart from MPA, SiC has a strong $\chi_2$ nonlinearity due to its non-centrosymmetric crystal structure. As a result, it is also suitable for efficient second harmonic generation (SHG). Fig 2(b) shows the intensity dependence of back scattered SHG from SiC. The slopes of the signals at the modulation frequencies of Δυ and 2Δυ in the log-log plot are 2.1 ± 0.1 and 2.2 ± 0.1, respectively, which are close to the values expected for second order nonlinear signal. Out of the two signals, the signal at 2Δυ can be used to approximately measure the intensity autocorrelation (see Appendix for the derivation). Similarly, the photocurrent signal at 4Δυ gives higher-order correlation function. The autocorrelation functions are shown in Fig 3(b). The autocorrelation trace obtained from the SHG has a width of 245 fs (full-width at half maximum), from which we estimate the duration of laser pulses to be about 175 fs assuming a Gaussian pulse profile. The trace from the photocurrent signal has a width of only 125 fs. While higher-correlation functions are expected to have widths that are shorter than intensity autocorrelation, it is interesting to note that correlation from 4PA actually is shorter than the laser pulse itself. This is an important observation that demonstrates the possibility of sub-pulse ultrafast optoelectronic switching using high-order MPA.

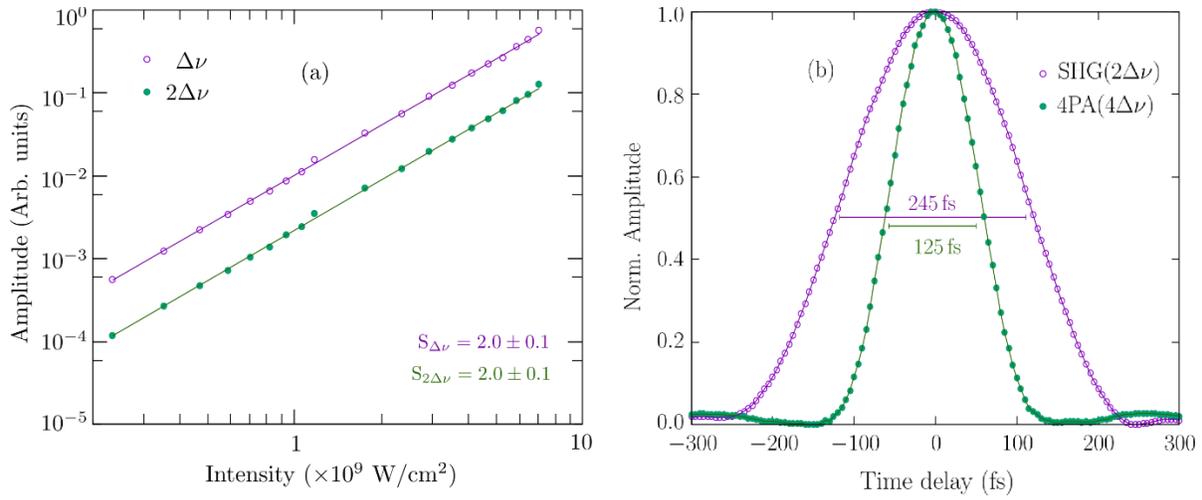

Figure 3. (a) Power-dependent photocurrent signals from SiC photodetector at modulated frequency Δυ and 2Δυ using 515nm. (b) Intensity autocorrelation obtained from the SHG modulated at 2Δυ and correlation trance obtained from the 4PA modulated at 4Δυ.

Finally, we discuss factors other than electronic structure that affect high-order MPA in SiC. The electronic structure of 4H-SiC is shown in Fig. 4. Three transitions are relevant for discussion here. First is the indirect transition from the valence band (VB) maxima at Γ to the M valley minima of the conduction band (CB). A minimum of three-photons in conjunction with phonons are necessary for this transition. Another indirect transition is from VB maxima to L valley minima with a bandgap of about 4 eV that requires 4PA assisted by phonons. The Γ-Γ valley direct transition with a bandgap of 4.8 eV requires four photons. Our results indicate that 3PA, although energetically allowed, does not contribute significantly in photocurrent excited by photons at 1.2 eV (1030 nm). The main contribution is from 4PA, which could include the Γ-L indirect transition and Γ-Γ direct transition. However, an indirect transition, because of momentum mismatch, is weak compared to a direct transition of the same order of MPA. Thus, only 4PA from the Γ-Γ transition has the dominant contribution to photocurrent in SiC. Note that this does not apply to all indirect bandgap semiconductors as previous measurements in GaP have shown that 2 and 3PA can have comparable contributions, where 2PA is an indirect and 3PA is a direct transition[27]. Why is MPA in SiC different? Here, we need to take into account the intensity at which the measurements are performed. In fact, Eq. (2) predicts a certain excitation intensity, also known as the critical intensity, above which the contribution from $(N + 1)$-PA become larger than that from $N$-PA[33]. The critical intensity depends on bandgap of a semiconductor: $I_c \propto E_g^4$, and in direct wide bandgap semiconductors with $E_g > 3$ eV, it is estimated to be in the order of $10^6$ GW/cm². As $I_c$ is higher than the excitation regime at which perturbative description of light-matter interaction can be applied (i.e. $I < 10^{12}$ W/cm²), it is not physically realized in the direct wide bandgap semiconductors. However, at some photon energies, photocurrents from $N$ and $(N + 1)$-PA can be comparable if $N$-PA is an indirect and $(N + 1)$-PA a direct transition. In GaP, the two contributions are comparable at the excitation intensities of few GW/cm². We expect comparable 3 and 4PA in SiC at similar excitation. However, such intensities are more than an order of magnitude lower than the lowest intensity of 68 GW/cm² at which we have been able to record photocurrents. Thus, the dominance 4PA in SiC is not solely due to the electronic structure of the material but also due to the photon density that is required to observe photocurrents. This may be true in other wide bandgap indirect semiconductors when excitation intensities are greater than a few tens of GW/cm². Nevertheless, SiC is special as it withstands such high intensities without showing saturation effects or photodamage.

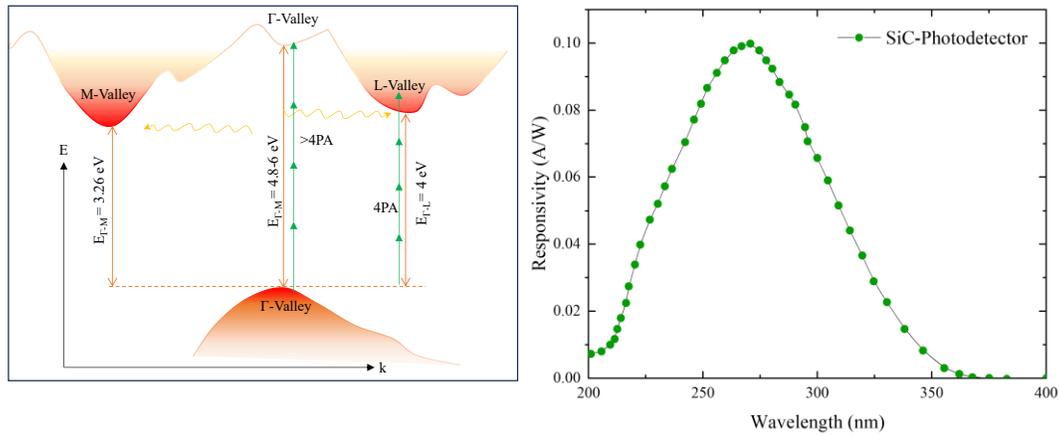

Figure 4. Schematic demonstrating after irradiation of femtosecond phase modulated pulses band structure of 4H-SiC. It has the highest occupied valance band at Γ-point and possible conduction bands at M, Γ, and L valleys with energy gaps of 3.26eV, 4.8eV, and 4eV, respectively. Γ- Γ or Γ-L transitions have four-photon absorption to excite electrons, reflecting a photodetector's peak responsivity at 275nm.

**Conclusion**

We have measured multiphoton photocurrent in 4H-SiC photodetectors at excitation intensities of 70—200 GW/cm$^2$ using phase-modulated femtosecond laser pulses at 1030 nm. Measurements show dominant contribution from the direct Γ- Γ transition by 4PA, while lower-order MPA which could result from indirect Γ-L and Γ- M transitions, trap related sub-bandgap absorptions are negligible. Lack photocurrent from sub-bandgap transitions makes SiC a good candidate for UV and MPA photodetector. Measurement of the response time of 4PA shows that it is shorter than the laser pulse itself indicating SiC is suitable for sub-pulse width switching in opto-electronics. Importantly, SiC can sustain excitation intensities up to 200 GW/cm$^2$, without noticeable saturation effects or photodamage making it an ideal system for high-power nonlinear optoelectronics.

**Appendix**

The electric field due to two phase modulated pulses with an inter-pulse time delay $\tau$ can be written as

$$E(t) = A(t)\cos((v + p_1)t) + A(t - \tau)\cos((v + p_2)(t - \tau)),$$

where $A(t)$ is the pulse envelope. The intensity of SHG recorded by a detector is given by

$$I_{SHG}(t,\tau) \propto E(t)^4.$$

$I_{SHG}(t,\tau)$ contains terms that oscillate with optical frequencies, mainly $v, 2v, 3v$ and $4v$, and radio frequencies, mainly $\Delta v$ and $2\Delta v$. As the detector is not fast enough to respond at optical frequencies, only the terms modulated at radio frequencies can be observed in the photocurrent. The term at $2\Delta v$ is relevant for intensity autocorrelation. It is approximately given by

$$I_{SHG,2\Delta v}(t,\tau) \propto A^2(t)A^2(t-\tau)\cos(2\Delta vt - 2v\tau).$$

The signal recorded by the detector is given by

$$S_{SHG}(\tau) \propto \int_{-t_R}^{t_R} A^2(t)A^2(t-\tau)\cos(2\Delta vt - 2v\tau)\,dt,$$

where $t_R$ is the response time of the detector. Note that in the experiments, the pulse envelope is nonzero only at sub-picosecond time scale and $\Delta v$ is in the range of kHz, thus $\cos(2\Delta vt - 2v\tau)$ can be considered to be a constant in the integral. Its effect are observed only at macroscopic time (or time at which signal is recorded by the digitizer). Thus, the expression for the recorded signal can be approximated by

$$S_{SHG}(T,\tau) \approx \cos(2\Delta vT - 2v\tau)\int_{-t_R}^{t_R} A^2(t)A^2(t-\tau)\,dt$$

$$= \cos(2\Delta vT - 2v\tau)\int_{-t_R}^{t_R} I(t)I(t-\tau)\,dt,$$

where $T$ is the macroscopic time. Thus, the amplitude of the measured photocurrent signal at $2\Delta v$ gives the intensity autocorrelation.

**Acknowledgements**

KJK acknowledges financial support from the Guangdong Province Science and Technology Major Project: Future functional materials under extreme conditions (FMUXC) -